# Measuring the Importance of User-Generated Content to Search Engines

**Nicholas Vincent, Isaac Johnson, Patrick Sheehan, and Brent Hecht**

Northwestern University, Evanston, IL

{nickvincent, isaacj, PatrickSheehan2018}@u.northwestern.edu, bhecht@northwestern.edu

## Abstract

Search engines are some of the most popular and profitable intelligent technologies in existence. Recent research, however, has suggested that search engines may be surprisingly dependent on user-created content like Wikipedia articles to address user information needs. In this paper, we perform a rigorous audit of the extent to which Google leverages Wikipedia and other user-generated content to respond to queries. Analyzing results for six types of important queries (e.g. most popular, trending, expensive advertising), we observe that Wikipedia appears in over 80% of results pages for some query types and is by far the most prevalent individual content source across all query types. More generally, our results provide empirical information to inform a nascent but rapidly-growing debate surrounding a highly-consequential question: Do users provide enough value to intelligent technologies that they should receive more of the economic benefits from intelligent technologies?

## Introduction

Search engines are immensely popular and enormously valuable intelligent technologies. Over 92% of American adults use web search (Purcell, Brenner, and Rainie 2012) and Google.com is the most-visited website in the entire world (Alexa.com 2018). Moreover, Google makes over $20 billion per year from search advertising revenue (Townsend 2017) and Google's market capitalization is one of the highest in the world (Forbes 2018).

However, very recent work has suggested that search engines, despite their power and profitability, may be surprisingly dependent on a resource that is both volunteer-created and freely available: user-generated content (UGC), and specifically Wikipedia. In particular, McMahon et al. (2017) found that search engine result page (SERP) click-through rates dropped drastically from 26% to 14% when Wikipedia results were removed from SERPs. This drop in click-through rate – a critical search engine evaluation metric – is enough to easily wipe out gains made by even major improvements to search engine algorithms, for instance the introduction of deep learning (Clark 2015).

While McMahon et al. showed that Google search users have a strong preference for Wikipedia pages when they are surfaced, McMahon et al.'s study design did not allow them to ask an equally important question: *How often do search engines surface Wikipedia links – let alone links to other types of user-generated content – in the first place?* In other words, it is unclear how often users are able to act on their strong Wikipedia preference. This means the full real-world impact of McMahon et al.'s findings are also unclear.

In this paper, we perform a rigorous audit of Google's search engine to understand the extent to which Google surfaces links to English Wikipedia and other UGC. Specifically, we examined results across six categories of high-value queries selected for popularity, potential for advertising revenue generation, and potential to influence users' lives. Using software we developed and are releasing with this paper, we also robustly address potential confounding effects from geographic personalization, known to be by far the major source of variation in search results (Hannak et al. 2013; Kliman-Silver et al. 2015; Xing et al. 2014).

Our results both complement and strengthen the findings of McMahon et al. We find that across all six categories of important queries, Google is highly reliant on Wikipedia to perform its core mission of satisfying user information needs. For some categories of queries (*trending* queries and *controversial* queries), Wikipedia articles appear in over 80% of (first) results pages and appear in the particularly important "top three links" over 50% of the time. Even for types of important queries for which Wikipedia appears less often (e.g. some high-revenue queries), Wikipedia still appears in over 20% of results pages. More generally, Wikipedia was by far the single most prevalent source of links across all query types. In other words, in our study, Google returned links to English Wikipedia far more often than it did for any other website in the world.

We do also find, however, that the value of UGC to Google more or less stops with Wikipedia. While Google frequently surfaces content from platforms that the literature commonly considers to be UGC (e.g. social media platforms), our findings showed that most of this content comes from professional sources (e.g. corporations, journalists) rather than individual users. For instance, although tweets frequently appeared in SERPs in our study, these tweets were almost always from corporate accounts or official political accounts like those of U.S. senators.

Our results have important implications for a number of specific constituencies. Most notably, our Wikipedia findings raise the stakes of the large social computing and computational social science literatures on Wikipedia. Our results suggest that the findings in these literatures – e.g. those about gender biases (Hill and Shaw 2013; Wagner et al. 2015) and geographic content biases (Johnson et al. 2016; Hecht and Gergle 2009)– not only have an impact within the Wikipedia web site, but also affect popular search engines.

More generally, our Wikipedia findings contribute to a growing discussion (e.g. Hecht 2017; McMahon, Johnson, and Hecht 2017; Lanier 2014; Posner and Weyl 2018; Vincent, Hecht, and Sen 2019) about the relationships between end users and intelligent technologies like search engines. Our results – along with those of McMahon et al. and others – highlight that end users are not just silent consumers of powerful intelligent technologies. Rather, through the content that they create, end users play an absolutely critical role in helping these technologies accomplish their core goals. This critical role is the basis for the nascent-but-burgeoning debate about the current distribution of financial rewards from intelligent technologies, a debate to which our results provide valuable early empirical information.

## Related Work

### Web Search and User-Generated Content

This paper was directly inspired by McMahon et al.'s work showing that Wikipedia is critically important to the success of web search (McMahon, Johnson, and Hecht 2017). As is discussed above, our research fills in a key piece of the puzzle outlined by McMahon et al. by examining how often Wikipedia results appear on Google SERPs. Our paper is additionally motivated by recent work by Vincent et al. (2018). Focusing on the relationships between Stack Overflow, Reddit, and Wikipedia, Vincent et al. found a similar – though smaller – effect for the amount of value Wikipedia adds to these external sites. It is important to note that McMahon et al.'s and Vincent et al.'s research was itself directly motivated by a call from the Wikimedia Foundation (the operator of Wikipedia) for research into the relationships between Wikipedia and its broader ecosystem, including search engines (Taraborelli 2015).

Researchers from tourism studies and medicine have also investigated the important role UGC plays in serving domain-specific search queries. In 2010, Xiang and Gretzel (2010) conducted an audit study using tourism-specific queries and found that social media platforms like TripAdvisor and Yelp made up about 11% of all Google results they collected. Haiyan (2010) performed a very similar study using the Baidu search engine and Chinese tourism queries and found social media comprised almost 50% of results. Laurent and Vickers (2009) studied the role of Wikipedia in serving health queries and found that Wikipedia was common: in their study Wikipedia appeared in 71-85% of top-ten results across multiple health-related query sets. Interestingly, this statistic is quite a bit higher than we observe for medical queries, a point to which we return below.

Outside of the academic literature, the search engine optimization (SEO) industry has leveraged the role of UGC in search ranking algorithms (Klais 2010; Zadro 2014). In fact, SEO firms are known to manipulate UGC (e.g. editing Wikipedia pages) to attempt to boost the rank of webpages of their clients (Shivar 2017).

### Search Engine Personalization Auditing

Many of our methodological choices below draw heavily from the findings and best practices of work that has sought to audit the degree of personalization in web search. It has been consistently shown in this literature that location is an important driver of personalization in search results. For instance, examining search personalization with respect to a multitude of factors such as gender, age, education, and browser choice, Hannak et al. (2013) found geographic location to be the main source of personalization (outside logging into a personal account; see below). Similarly, in a user-focused study of search personalization, Xing et al. (2014) found evidence of substantial personalization due to location. When doing more focused analyses on the role of geographic location in search personalization, Kliman-Silver and colleagues (2015) found that the magnitude of geographic personalization varied with query type. This is one reason why we examine six types of queries in this work rather than focusing on a single type.

In our Methods section below, we discuss in significantly more detail how the personalization auditing literature inspired and informed our methods.

## Methods

In this section, we describe the five key aspects of our methodological approach: (1) our software framework, (2) how we selected queries, (3) how we analyzed SERPs, (4) how we identified UGC, and (5) how we handled the potential confound of geographic personalization.

## Software Framework

The high-level methodological challenge faced in this research was to collect Google SERPs for many queries from a variety of simulated locations. To address this challenge, we built a software package that modifies and extends the open-source, Selenium-based SerpScrap (Schmidt 2018) library, which automates the desktop version of Chrome web browser. In this paper, we focus on desktop search and leave to future work extending our analyses to incorporate the nuances of mobile search (see Discussion below). We make our software available with this paper to allow others to re-purpose and/or replicate our approach[1]. We note that utilizing the software will require moderate updates due to the constantly changing structure of Google SERPs.

Our software iterates through queries (selected as described below) and locations (also selected as described below) in quick succession but pauses for a full minute between each query to avoid causing undue load. While this approach is inspired by past work by Kliman-Silver and colleagues (2015), it also differs from this work in one key way: Kliman-Silver and colleagues took samples for a single query at one time instant, whereas we issue queries sequentially. We believe our sequential approach, which reduces the resources required to collect data, is appropriate because a conclusion of Kliman-Silver et al.'s work was that personalization is consistent over time. Indeed, we were able to verify that our sequential approach led to similar levels of personalization as Kliman-Silver et al.'s parallel approach: our results replicate the general levels of personalization found in their work (our SERPs had an average Jaccard Index of 0.86 and an average edit distance of 1.9, within the range of values observed by Kliman-Silver et al.).

To simulate queries from different locations, we also take inspiration from Kliman-Silver et al. (2015). Specifically, following their approach, we inject Javascript that overrides the *geolocation.getCurrentPosition*() function to return a latitude and longitude of our choice. We then automatically click the "update location" button and refresh the SERP. We verified that this approach worked as it did in Kliman-Silver et al. by leveraging the fact that Google reveals the perceived location of each query at the bottom of each SERP. For instance, a query from Chicago will have the following text at the bottom of the resulting SERP: "Chicago, IL. Reported by this computer".

Our software only captures the first SERP for each query. We focused on the first SERP as research has shown that users only very rarely look at results pages beyond the first one (Van Deursen and Van Dijk 2009). For a similar reason, in our analyses, we provide an additional level of focus on the top three results on the first SERP. Previous research shows the higher-ranking positions in search results are more valuable (Radlinski and Joachims 2005) - the first spot may receive up to 30% of all traffic, with the top three spots receiving 60% of all traffic (Insights 2013).

## Selecting Queries

In the search literature – and certainly in the search auditing literature – deciding on a set of queries for an analysis is well-known to be challenging (Pan et al. 2007; McMahon, Johnson, and Hecht 2017; Hannak et al. 2013; Kliman-Silver et al. 2015). Aside from researchers operating within search companies (and sometimes even for these researchers), it is impossible to obtain a set of queries that is guaranteed to be representative. As a result, researchers must use heuristic strategies to generate an imperfect query sample that still can provide insight for their research questions. A typical approach involves first choosing a limited set of query types that have significant real-world implications, e.g. queries related to medical issues, commerce, or politics (Pan et al. 2007; Kliman-Silver et al. 2015; Epstein and Robertson 2015; Hannak et al. 2014; Kulshrestha et al. 2017; Xiang and Gretzel 2010; Laurent and Vickers 2009). By focusing on a single or small set of query types, researchers can then use creative approaches to generate viable specific queries within these type(s), e.g. manually adapting a query dataset from a published paper to a new geographic context (McMahon, Johnson, and Hecht 2017; Haiyan 2010), using externally available data to generate potential queries (e.g. Laurent and Vickers 2009; Hannak et al. 2013), or manually generating reasonable queries (e.g. Xiang and Gretzel 2010; Kliman-Silver et al. 2015).

In our research, we sought to adopt a diversified version of the above approach by including six separate query types instead of just one or two. More specifically, we focused on query types with real-world importance along three dimensions: (1) how often a query is made (popular queries), (2) the revenue Google makes from selling ads on the query (high-revenue queries), and (3) the degree to which the results of a query could impact users' lives (influential queries). In keeping with approaches in the search literature, we used a combination of external resources such as Google Trends and Google AdWords and data from existing research (Hannak et al. 2013; Kliman-Silver et al. 2015) to select queries in a systematic way. For each of the three dimensions above, we developed two separate categories of queries, leading to six total query categories. Each query category contains between 10-20 queries, a number selected to be practical with respect to the rate limit we imposed to avoid excessive querying.

---

[1]https://github.com/nickmvincent/SerpScrap for data collection code; https://github.com/nickmvincent/you-geo-see for analysis code and data

By considering three different dimensions of importance and using two different categories of queries for each dimension, our intention was to gain a broad and robust view of the role that UGC plays in Google SERPs, and one that is not unduly contingent on query-specific idiosyncrasies. Below, we detail our categories and their constituent queries. For replication and extension purposes, a full list of our queries and their assigned categories can be found in our software repository linked above.

**Popular Queries:** We considered two categories of popular queries: *trending* queries and *most-popular* queries. To develop our *trending* category of queries we turned to Google Trends. Google Trends is a public website that Google maintains to share data about patterns in usage of Google's search engine. We used Google Trends' "trending searches" feature to obtain queries characterized by a large baseline number of searches and a spike in searches (typically 1,000,000+ searches) ("Google Trends" 2018). Specifically, our *trending* category consists of each daily top trending query from Nov. 28 to Dec. 7, 2017 (10 queries).

With respect to our *most-popular* category, Google Trends does not directly provide a list of the most popular queries on Google's site overall, but it does do so for specific query topics. In other words, we can know what queries are popular within a topic, but we do not know the global popularity of the topic. As such, to develop our *most-popular* query category, we collected the top three queries by U.S. query volume across a set of Google Trends-defined query topics which were commercial or political in nature: Auto Companies, Fast Food Restaurants, Financial Companies, Governmental Bodies, Politicians, and Retail Companies. We discuss the potential limitations of manually selecting categories from Google Trends' offerings in our Limitations section.

**High-Revenue Queries:** Google sells many of its ads – and generates much of its revenue (Shaban 2018) – by allowing entities to bid on SERP ad placements on a query-by-query basis (using a system called Google AdWords). While Google does not provide high-level data about which are the most expensive queries, the SEO industry has published informal studies on this topic. According to one such study, *insurance*-related queries and *loan*-related queries are two of the most expensive categories of queries (wordstream.com 2011) and, as such, we selected these two categories to represent high-revenue queries. To populate these categories with actual queries, we used Google Trends' "Explore" feature to obtain the top ten queries for "insurance" and for "loans" (in the U.S., from all of 2017). We used Google AdWords' Keyword Planner to verify that the bids for these query categories were indeed very high; we observed a top cost-per-click of $514 for the most expensive query in the *insurance* category and $259 for the most expensive query in the *loans* category in December 2017.

**Influential Queries:** Query popularity and query revenue do not necessarily correlate strongly with the influence of a SERP on people's lives. Some types of queries – e.g. queries related to a family member's serious illness or queries related to informing one's political views – can have an outsized impact (Epstein and Robertson 2015; Soldaini et al. 2016). To gain a sense of UGC's influence in Google search results for particularly influential queries, we included two additional categories of queries that have been the subject of prior research in the search literature because of their influential nature (Kliman-Silver et al. 2015; Epstein and Robertson 2015; Soldaini et al. 2016): queries on medical topics and controversial topics. For our medical category, we use a subset of queries from Soldaini et al.'s study of health searches on the Bing search engine (Soldaini et al. 2016). This set consists of 50 queries sampled from Bing's top 500 medical queries; we used the first 20 queries. For our controversial query category, we were unable to re-use queries from Epstein's experiment or Kliman-Silver's audit study because the queries were related to current events (e.g. topics included the UK Prime Minister Election, Barack Obama's US presidency). As such, to systematically generate a diverse list of up-to-date search queries, we used the top ten topics from procon.org, a non-profit organization that hosts information about controversial issues.

## Understanding SERPs

Modern Google SERPs consist of substantially more than the traditional "ten blue links" (Chen et al. 2012) that formally comprised the canonical search results page. Current SERPs contain multiple columns of content, and items like carousels (which have multiple links per row), answer boxes, and more. To understand the prominence of UGC on Google SERPs, it was important that we account for all this complexity.

As such, in addition to standard "blue links", our analytical framework also explicitly considers the following Google SERP element types, which are also visualized in the diagram in Figure 1:

- *NewsCarousel*: A row of three cards that each link to a news story.
- *TweetCarousel*: A row of three cards with one tweet each. Google obtains the tweets either from Twitter's search (a *SearchTweetCarousel*) or a single user (a *UserTweetCarousel*).
- *MapsBox*: A box with Google Maps embedded that includes up to three locations. We mainly observed *LocationsMapsBox* elements, which have entries corresponding to multiple locations of a single entity (i.e. business).
- *AnswerBox*: A box that includes a link to a website and a snippet of text meant to answer a question; includes variants such as *PeopleAlsoAskAnswerBox* elements.

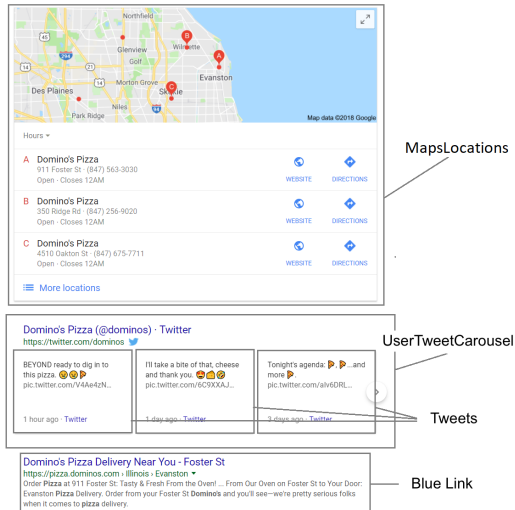

Figure 1. A screenshot depicting a selection of elements on Google SERPs.

It is important to note that for most of our analyses below, we do not consider links that occur in the "*Knowledge-Panel*", another SERP element that, for desktop web browsers, appears on the right-hand side of SERPs. Although KnowledgePanels include Wikipedia and social media links, we did not consider them in our core analyses because content in KnowledgePanels cannot be easily assigned a rank (as the panel essentially exists separately from the ranked search results). However, we did perform a small analysis of how our results might change if we did consider KnowledgePanels, and we discuss that analysis below. With respect to implementation, to operationalize our above framework (e.g. to store a link that appears as a blue link in our database as such), our software parses the CSS (cascading stylesheets) associated with each SERP. Since elements are represented the same away across SERPs, this is a straightforward task.

**Metrics**: Given the complexity of SERPs, there are many metrics one could use to understand the prominence of UGC in SERPs. The primary metric on which we focus is the *incidence rate* of each domain and element, an approach that is typical when assessing the prominence of content on SERPs (Xiang and Gretzel 2010; Haiyan 2010; Laurent and Vickers 2009; Robertson, Lazer, and Wilson 2018). The incidence rate is the fraction of SERPs in a given query category on which a domain or element appears. For instance, if a Wikipedia link appears in 10% of SERPs for a given query category, Wikipedia would have an incidence rate of 0.10 for that category.

As we discussed above, research has shown that higher-ranking content gets substantially more traffic, so we also calculate the *top-three incidence rate* of each domain or element. This is the fraction of SERPs that have a given domain or element in their top three rows. When necessary to avoid ambiguity between incidence rate types, we refer to the basic form of incidence rates as *full-page incidence rates*.

In calculating both incidence rates, we treat SERP elements like carousels as a single item (hence the NewsCarousel has its own incidence rate), but also count the content of carousels as items (i.e. the tweets and news articles). For example, if there is a SERP with a *New York Times* article embedded in a rank 2 NewsCarousel, then that SERP will increase the top-three incidence rate for the domain ny-times.com *and* the element NewsCarousel.

## Classifying Content as UGC

A critical requirement of our analyses is the ability to distinguish between UGC and non-UGC search results. Because there is no consensus definition of UGC (Vickery and Wunsch-Vincent 2007), we operationalized two definitions from the literature: a *platform-centric* definition and a *content-centric* definition.

The platform-centric definition is one that is commonly used in UGC research, often implicitly (e.g. Cha et al. 2007; Jin, Phua, and Lee 2015; Latham, Butzer, and Brown 2008; Hecht and Stephens 2014). Broadly speaking, this definition assumes that any content is UGC if it appears on a platform that hosts a large amount of UGC. There are a few prominent definitions of UGC that explicitly adopt a platform-centric perspective. For instance, Luca (2015) provides a categorized list of popular UGC platforms and Dhar and Chang (2009) define UGC as the "conjunction of blogs and social networking sites".

To operationalize our platform-specific definition, we cross-referenced the list of domains encountered in our data collection process with those on Luca's 2015 list. More specifically, this means that, under the platform-centric definition, we categorized as UGC any content from the following domains: Wikipedia, Facebook, Twitter, YouTube, Instagram, Yelp, LinkedIn, and TripAdvisor. The coding we describe here was applied only to content from these domains.

In a 2007 report (Vickery and Wunsch-Vincent 2007), the Organization for Economic Cooperation and Development (OECD) offered a stricter definition of UGC that focused on content rather than platform. Under the OECD definition, UGC must (1) be published, (2) require some creative effort (i.e. not be a copy of some existing content), and (3) be created "outside professional routines and practices." To determine whether content was UGC under this definition, we used a qualitative coding approach and assessed whether a search result (e.g. a tweet in a Twitter carousel) met criteria #2 and #3 (all results implicitly met criteria #1). Specifically, our codebook instructed coders to view each search result (both the content on the SERP as well as the content on the linked website, e.g. a Twitter page) from our list of UGC platforms and identify (1) if the content appeared to be

"creative" (i.e. not a copy of some other content) and (2) if the content appeared to be authored outside of professional "routines and practices". Coders used contextual information such as Twitter biographies or the presence of user reviews to judge whether the content appeared to be "professional" or not.

As we describe below in Results ("Types of UGC"), our results highlighted an interesting and meaningful contrast between the platform-centric and content-centric definitions of UGC. Upon discovering this contrast, we sought to better understand it by classifying content along two additional criteria that we hypothesized would be insightful based on what we saw in exploratory analyses. The first axis was related to who appeared to have authored the content: either an individual, an organization, or a bot. The second axis was related to the type of actor that authored the content. Using an inductive approach, we identified four types of individuals (*journalist, political figure, celebrity, other*) and five types of organizations (*journalistic*, *political*, *corporate*, *non-profit*, *other*).

To assess the reliability of the full coding scheme, we sampled up to 10 items for each UGC domain from our first dataset (a comparison of urban and rural search results: see below) and two researchers coded these samples. The researchers achieved substantial or perfect agreement in every case. For Facebook, Instagram, LinkedIn, Yelp, and TripAdvisor the coders achieved perfect agreement. They achieved a Cohen's Kappa of 0.87 and 0.67 for YouTube and Twitter respectively. Given this level of agreement, only one researcher coded the remaining samples.

## Controlling for Geography

As noted above, based on our review of the search personalization literature, we expected that the importance of Wikipedia and UGC to Google might vary extensively by query location. As such, we developed a rigorous infrastructure to issue queries from a variety of carefully chosen simulated locations and planned to report our results with ranges defined by our location-specific results. However, upon running our experiments with this framework, we found that with respect to the prominence of Wikipedia and UGC in Google SERPs, there was little geographic variation.

As such, we in fact ran two separate experiments in this research project: (1) an experiment that rigorously considered potential geographic variation and, upon finding that this did not exist, (2) an experiment that used a simple population-weighted sample of geographic queries that could provide reliable single results for metrics of interest instead of ranges. We describe the methods we used in each of these experiments in turn below. The results of each experiment are discussed in detail in the sections that follow, as are the implications of the lack of variation that we observed.

**Geographic Variation Experiment:** Our geographic variation experiment was rooted in a spatial sampling approach that was designed to understand the maximum variation of UGC incidence across geography while at the same time maintaining a reasonable query rate. Our sampling strategy targeted three spectra on which the behavior of intelligent geographic technologies are known to vary: rural/urban, socioeconomic status (SES), and political preference (e.g. Johnson et al. 2017; Hecht and Stephens 2014; Cohen and Ruths 2013).

As our researchers had the most familiarity with Google's US search results, we restricted our focus to the United States. Choosing a specific study area for these and similar reasons is a common design choice in "GeoHCI" (Hecht et al. 2013) and computational social science (as well as many other fields) (e.g. Hecht and Stephens 2014; Mahmud, Nichols, and Drews 2012; Jurgens et al. 2015). We discuss possible expansions of this work to different geographic contexts in Future Work.

To generate specific geographic coordinates for the strategy outlined above, we used the following approach:

1. Urban-Rural: Using the urban-rural classifications by the U.S. National Center for Health Statistics (NCHS) (Ingram and Franco 2014), we sampled 10 counties from the most urban and most rural classes. These NCHS classifications are often leveraged in GeoHCI examining rural-urban issues (Colley et al. 2017; Thebault-Spieker, Hecht, and Terveen 2018; Johnson et al. 2016). We then used the centroid latitude and longitude provided by the U.S. Census for each county as a query location.

2. Income: We selected the top and bottom 10 counties in terms of 2015 median income, according to the 2011-2015 U.S. Census American Community Survey 5-Year Estimates (U.S. Census Bureau 2011), and executed the county-to-coordinate mapping as described above.

3. Voting: We selected the top and bottom 10 counties in terms of percentage of votes for Hillary Clinton in the 2016 U.S. Presidential election and again executed the same county-to-coordinate mapping. This county-level data was published by Townhall (Townhall.com 2017) and accessed via McGovern's repository (2017).

**Population-weighted Experiment:** As reported below in Results, the rigorous geographic comparisons described above showed little evidence of geographic variation in metrics of interest. As such, it was reasonable to use a single set of query locations to report our results. However, it was non-optimal to select one of our pre-existing location sets (e.g. most-rural or wealthiest) as representative and report those results for two reasons: (1) we did observe a (quite) small amount of variation across the spectra outlined above and (2) doing so may raise other ecological validity concerns.

As such, in our second experiment, we developed a new set of query locations that avoided both of these issues. To develop this set, we randomly sampled 40 U.S. counties using a population-weighted approach. We then used the U.S.

Census-provided representative coordinate for each of these counties as our query locations. We issued each query from each of these coordinates and report results averaged across all coordinates. Experiments were run in January 2018.

## Results

We first report the results of our population-weighted experiment described above. We then provide additional context with the results of our geographic variation experiment.

## Population-Weighted Experiment

Figure 2 summarizes the results from our population-weighted SERP dataset. The figure shows the full-page and top-three incidence rates for all UGC domains in the dataset, as well as the top five non-UGC domains and SERP elements to provide context. Table 1 zooms in and focuses on the results for Wikipedia specifically.

The strongest signal present in both Figure 2 and Table 1 is that Wikipedia is absolutely critical in Google's approach to responding to queries. More specifically, Figure 2 and Table 1 show that Wikipedia is not only the most prominent

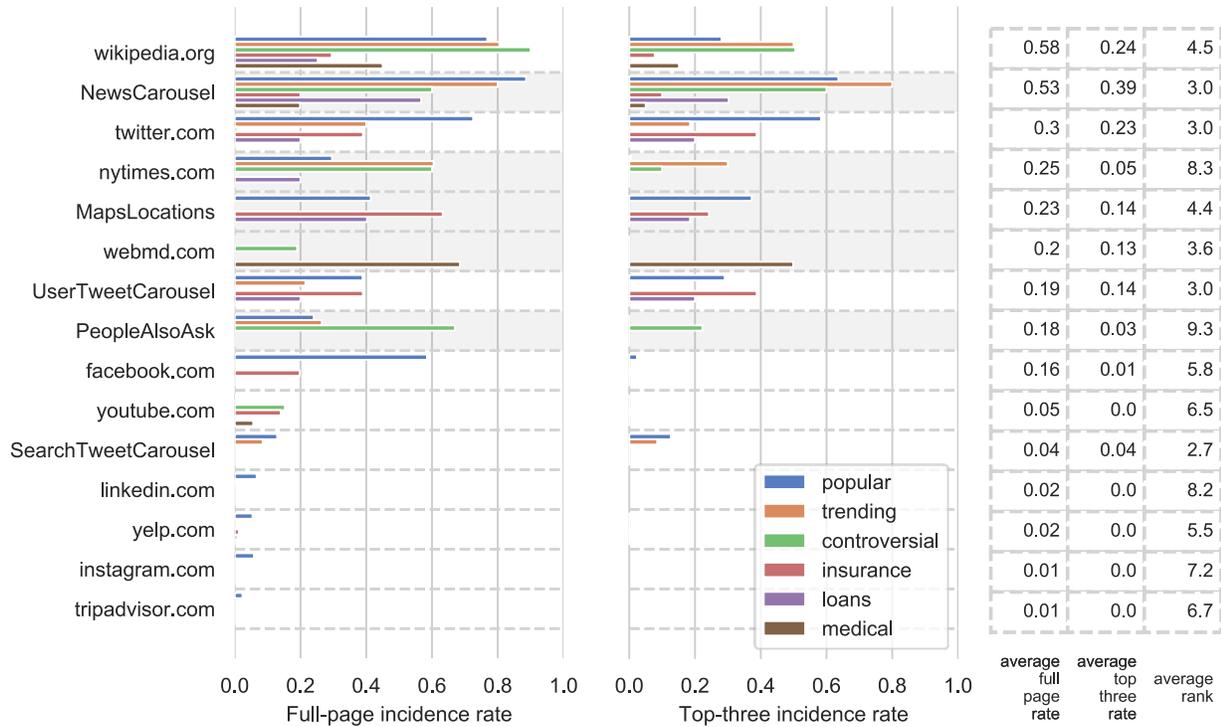

Figure 2. This figure summarizes key metrics for all UGC domains in our study and the top 5 non-UGC domains/elements (highlighted in light grey). Rows are ranked by average full-page incidence rate, shown on the right, followed by average top three incidence rate and average rank for each domain. The average incidence rates should be interpreted with a degree of caution as they are not intended to be representative of overall incidence rates, just the average across the six query categories we analyzed in this study.

| Query Category | Top Domain *The top domain in terms of full-page incidence rate.* | Rank of Wikipedia in Incidence Rate *i.e. if Wikipedia is top domain, this equals 1.* | Wikipedia Full-Page Incidence Rate *The fraction of SERPs with links to Wikipedia* | Wikipedia Top-3 Incidence Rate *The ratio of top-3 SERP results with links to Wikipedia* |
|---|---|---|---|---|
| Trending | Wikipedia | 1 | 0.81 | 0.50 |
| Most-Popular | Wikipedia | 1 | 0.77 | 0.28 |
| Loans | wellsfargo.com | 6 | 0.25 | 0 |
| Insurance | progressive.com | 10 | 0.29 | 0.08 |
| Controversial | Wikipedia | 1 | 0.90 | 0.51 |
| Medical | webmd.com | 5 | 0.45 | 0.15 |

Table 1. A targeted look at the prevalence of Wikipedia in Google SERPs by query category. Does not include SERP elements.

UGC platform on Google SERPs but also is in fact the single most prominent website of any kind on Google SERPs. For *trending* and *controversial* queries, Wikipedia appears in over 80% of first SERPs and rivals the prominence of structural elements like the NewsCarousel. For *insurance* and *loan* queries, the lowest incidence rates for Wikipedia in our study, Wikipedia still appears in over 25% of SERPs. When considering only the top-three result rows, Wikipedia remains very prominent, showing up in 50% of top-three results for some categories of queries (*trending* and *controversial*). In aggregate, the average full-page SERP incidence rate for Wikipedia in our study was 0.58 (58% of SERPs had Wikipedia pages). Twitter's 0.30 is the next-highest average full-page rate.

One concrete example of a query in our sample for which Wikipedia is very important is "minimum wage" from our *controversial* query category. The SERPs for "minimum wage" had a link to the Wikipedia article "Minimum wage in the United States" in two places: a rank 1 AnswerBox, and a blue link at rank 6. On the other hand, an example of a query for which Wikipedia is less important is the query "life insurance," where Wikipedia showed up at rank nine.

Beyond Wikipedia, Figure 1 additionally shows that Twitter is also important to Google's ability to respond to queries in many of our categories. For instance, for *most-popular* and *trending* queries, the full-page Twitter incidence rate is above 40% and the top-three incidence rate for *most-popular* queries is higher than that of Wikipedia. Interestingly, however, Twitter almost never appears in *controversial* and *medical* SERPs, perhaps a reflection of a specific design decision at Google.

**Types of UGC:** Our Twitter results – combined with the non-trivial prevalence of other UGC platforms for certain types of queries (e.g. Facebook for *most-popular* and YouTube for *controversial*) – seemingly suggest that Google's dependence on UGC extends significantly beyond Wikipedia. Indeed, using a platform-centric definition of UGC, this is the case.

However, the results of our content-centric qualitative coding exercise demonstrate that the platform-centric perspective is problematic for our research. Of the 345 unique social media results that appeared in our collected data, 95% failed to meet the OECD's content-centric definition of user-generated content (we note that together, these links appeared approximately 5,000 times in our dataset, because many results like the TweetCarousel and corporate social media pages were identical across locations). In particular, 73% of the links came from an official corporate account. Another 14% were from an official political account and 4% were journalistic. This means that while Google surfaces a substantial amount of content from non-Wikipedia UGC platforms, almost all of this content is not UGC from a content-centric perspective. Instead, this content resembles that on typical webpages: it is written by professionals. We return to this point in the Discussion.

**Effect of the Knowledge Panel:** While the Knowledge Panel lacks a "rank" in the desktop version of Google, it is still possible to re-compute the full-page incidence rate of each domain including the Knowledge Panel. Since Wikipedia is prominent in the Knowledge Panel, this calculation substantially boosts Wikipedia's average full-page incidence rate from 58% to 69%. Although we saw some social media links in the Knowledge Panel, every one of them linked to an organizational account, so these would not influence our conclusions above.

## Geographic Variation Experiment

As noted above, we were interested to find that – although prior work has highlighted the influence of location-based personalization for some queries – we saw very few meaningful differences in the importance of UGC across the spectra that we considered (urban vs. rural, SES, political preference). For the few cases in which we saw meaningful variation, the effect size was quite small.

We assessed the variation across the three geographic spectra by comparing full-page and top-three incidence rates from one end of each spectrum to the other. We performed these comparisons for every UGC domain (thus assuming a platform-centric definition of UGC) and across every query category. We tested to see if different types of locations saw different UGC domains at significantly different rates. To compute the median difference in incidence rates, we only considered the 115 comparisons in which a domain appeared at least once (e.g. Yelp pages never appeared for *medical* queries, so we did not include geographic comparisons of Yelp incidence rate for *medical* queries).

For Wikipedia, the median full-page incidence rate difference across all spectra and query categories was only 0.01, and the maximum was only 0.16 (political spectrum and *most-popular* queries); the Wikipedia top-three incidence rate median difference was 0 and the maximum was 0.07. The median across all 115 comparisons was 0.01. Moreover, of these comparisons, only 15 differences were statistically significant based on Fisher's exact test ($p < 0.05$), i.e. in most cases we fail to reject the null hypothesis that UGC is equally likely across geographic strata. When considering only links that meet the content-centric definition of UGC, only 9 differences were significant. In other words, observable geographic variation in UGC was rare.

## Discussion

### Distribution of the "Technological Dividend"

The most significant signal in our results is the critical role that Wikipedia plays in helping Google accomplish one of its most important goals: satisfying user information needs. For the English-language queries that we considered, Google is more dependent on Wikipedia than any other website in the world. Moreover, for some of Google's highest-volume queries (*trending* and *most-popular*), Wikipedia appears on a large majority of results pages.

These results help to inform a highly-consequential discussion about the economics of computing that is moving from the margins of the literature (e.g. Arrieta Ibarra et al. 2018; Hecht 2017) and into mainstream debate (e.g. Madsbjerg 2017; Posner and Weyl 2018; Porter 2018; Kugler 2018). This discussion centers on potential asymmetries in the relationship between users and lucrative intelligent technologies: user-generated data is immensely important to such technologies, but many argue that users are not receiving a proportional share of the economic benefits from these technologies. Our results certainly point to one such potential asymmetry: the Wikipedia community creates tremendous value for search engines, but search engines only donate a relatively small amount of money to support the Wikimedia community (Parr, Ben 2010; Seitz-Gruwell 2019). This finding raises provocative questions that can advance this discussion, e.g. are Wikipedia editors some of the most important and underpaid employees of search companies?

Arrieta Ibarra and colleagues (2018), Hecht (2017), and others (e.g. Posner and Weyl 2018; Porter 2018) have identified information imbalances between intelligent technology owners and data creators as a key mechanism for the current distribution of economic benefits of intelligent technologies. While the developers of intelligent technologies know many such technologies would struggle substantially without constant "data labor" by their users and others (e.g. Wikipedia editors), most people have very little understanding of the value of their data-generating labor. These authors have argued that the research community should therefore work to level the playing field by measuring and making people aware of the value their data brings to intelligent technologies (Hecht 2017). Our results make a contribution towards this goal, and also point to the importance of engaging in similar investigations in related domains (e.g. OpenStreetMap, Wikidata).

As McMahon et al. (McMahon, Johnson, and Hecht 2017), Posner and Weyl (2018), and others (e.g. Porter 2018) have noted, the discussion about the distribution of the technological dividend must also consider the value of the service that intelligent technologies "trade" for data-generating labor. After all, most Wikipedia editors benefit heavily from their use of Google, and McMahon et al. showed that Wikipedia itself does as well (McMahon, Johnson, and Hecht 2017). As such, our results point to an additional important area of future research: doing qualitative and quantitative work to understand whether the Wikipedia community believes anything should change in Wikipedia's relationship with intelligent technologies given the increasing informational equality on this topic.

Additionally, our results also highlight a related line of inquiry centered around a key question: How can we reduce any discrepancy between the value created by data like Wikipedia articles and the rewards received by those who created the data. Hecht (2017) and others (e.g. Arrieta Ibarra et al. 2018) have suggested that collective action by users – e.g. through boycotts, "data strikes", or data unions – can be one possible solution. Indeed, recent research has highlighted the potential impact that data strikes, boycotts, or combinations thereof could have on intelligent technologies (Vincent, Hecht, and Sen 2019). However, other, less confrontational approaches (which may also be more immediately tractable than widespread data strikes or data unions) may be possible and are likely desirable. For instance, just making visible the value of Wikipedia to search engines could encourage search engine companies to more prominently credit Wikipedia through design changes or to contribute donations of money or data.

More generally, Madsbjerg (2017) and McMahon et al. (2017) have argued that one major challenge in analyzing the value of user-provided data to profitable intelligent technologies is that the required datasets for such analyses are almost always private. Our paper – along with McMahon et al. – highlights an approach we believe can, at least initially, be quite effective at addressing this challenge: focusing on Wikipedia and other UGC rather than more difficult-to-access types of user-generated data such as search logs and personal information that also play critical roles in intelligent technologies. Our results show that just focusing on more open types of information is sufficient to at least begin the empirical examination of these issues.

## Wikipedia Matters outside Wikipedia

The social computing and computational social science communities have developed a large literature on Wikipedia. This literature has examined topics ranging from the content coverage biases that exist in Wikipedia (e.g. (Reagle and Rhue 2011; Johnson et al. 2016; Hecht and Gergle 2010; Pfeil, Zaphiris, and Ang 2006) to the collaboration patterns between editors that lead to the highest-quality content (e.g. Zhu, Kraut, and Kittur 2012; Zhu et al. 2013).

Our results further bolster the importance of this literature by showing that the literature's findings have implications far beyond the boundaries of Wikipedia. For instance, prior work has shown that the English Wikipedia has more missing articles about women than about men (Reagle and Rhue 2011) and similar patterns have been observed with respect to Wikipedia's coverage of some geographic areas versus others (Johnson et al. 2016). Our results highlight that not only do these biases affect reader experience on Wikipedia, they also affect Google's ability to address information needs associated with the disadvantaged topics. That is, if Wikipedia has less information about a topic of interest to a certain group, this will also affect Google's ability to address information needs related to this topic.

## Definitions of User-Generated Content

Our findings related to the platform-centric versus content-centric definitions of UGC may have important methodological implications for some UGC research. In particular, we found that the platform-centric definition of UGC was problematic in the context of our study. Had we relied on this definition exclusively, we would have believed that everyday Twitter users were powering Google at almost the same rate as Wikipedia editors. Instead, thanks to our qualitative analysis, we discovered that at the content-level, the vast majority of tweets surfaced by Google in our study were not UGC but rather were written by professionals. In other words, these tweets are analogous to short-form company websites (and, in some cases, news articles).

This result highlights calls (e.g. Ruths and Pfeffer 2014) for researchers to consider the nature of content on platforms that host UGC like Twitter before making assumptions about its professional or amateur nature. While much research currently takes care to do basic filtering for bot-created content – and there are well-known approaches for doing so in certain platforms (e.g. Davis et al. 2016) – filtering out organizational and other professional accounts will be more difficult and is deserving of further research along the lines of McCorriston et al. (2015).

## Geographic Personalization and UGC

Our geographic comparisons suggest that personalization based on geographic location may be non-substantial for certain types of search phenomena. This may simplify methods for some search auditing research projects, but more work is needed to understand when controlling for geography is necessary and when it is not. We note that we did see substantial variation for content other than UGC, e.g. Google Maps SERP elements. Additionally, given that Wikipedia is not equally comprehensive in all languages (Hecht and Gergle 2010) and that platforms like Twitter are not equally popular in all countries (Schoonderwoerd 2013), geography likely matters *across* national and linguistic borders. Future studies should address this directly.

## Limitations

As is typical in the search auditing literature, although we aimed to generate queries systemically, the immense number of search engine use cases makes it impossible to generate a truly representative query set for data collection (at least from a position outside of an institution that operates a large search engine). As such, we emphasize that our results must be considered in the context of the queries we selected.

This means that while our results likely generalize to many queries that are similar to our query sets, for instance queries about popular commercial entities or queries about common health problems, our results may not generalize to all search engine use cases, such as complex, infrequently-made queries. Furthermore, our *most-popular* query set is constrained by Google's willingness to share query volume data, as well as the manual process of selecting categories from Google Trends. It should also be noted that this query set may contain some thematic overlap (though no actual query overlap) with other query sets, e.g. *controversial.*

While we controlled for the effect of geography within the US, extending this analysis to include additional countries as noted above could provide valuable insight into the importance of Wikipedia and UGC globally. This analysis would require parsing other prominent search engines (e.g. Naver) and identifying appropriate geographic spectra.

Another direction for future research would to be to expand data collection and analysis to mobile devices. As mobile devices are used heavily for local search (Teevan et al. 2011), focusing heavily on local queries would make sense in this case. Given that we saw extensive variation across query categories, extending our work to consider additional categories would also be a useful direction of future work.

Search engines, like many intelligent technologies, are constantly changing. Therefore, longitudinal auditing will be valuable, both to account for revisions to SERPs – i.e. new specialty boxes and elements – and to account for algorithmic changes. For instance, the importance of UGC sources may vary as search engines integrate new techniques from the deep learning (e.g. RankBrain (Clark 2015) or structured knowledge domains (e.g. Knowledge Vault (Dong et al. 2014)). Indeed, the introduction of these technologies may be responsible for the decrease we observed in Wikipedia full-page incidence rate for medical queries relative to the work of Laurent and Vickers (2009) last decade (although the methods are not directly comparable). Doing this longitudinal auditing will require careful attention to edge cases, which means that recurring human validation (and likely updating our software accordingly) will be critical for future research in this direction.

To support such a longitudinal analysis, we are making our software available with the publication of this paper, although using this software will require updates based on changes to SERP structure. Researchers may also want to implement our approach using headless browser-based techniques, which likely require less overhead, or consider the recently released framework by Robertson et al. (2018) that uses on a Chrome plug-in and crowd workers from Crowd-Flower and Prolific Academic (Prolific 2018; Van Pelt and Sorokin 2012). Though Robertson et al.'s focus was on using their framework to study political personalization in web search, Robertson et al.'s data reveals they were able to replicate our results about the importance of Wikipedia in the political domain, adding an additional degree of rigor to the findings above.

Finally, UGC like Wikipedia has been used to train intelligent technologies, including by Google (e.g. for language understanding (Hewlett et al. 2016)). This is an entirely separate avenue by which UGC creates value for the owners of intelligent technologies. A very promising research direction would be to measure how the inclusion of UGC impacts the performance of these algorithms, similar to what Vincent et al. did in their recent work (Vincent, Hecht, and Sen 2019). However, doing the same for search engines and other private intelligent technologies will require creative approaches as it will likely require extensive access to proprietary software and data.

## Conclusion

This paper provides evidence that UGC, largely in the form of Wikipedia articles, is immensely valuable to web search engines. Examining six categories of queries, we found that Wikipedia's volunteers have created a resource that is critical to Google's ability to address information needs. Our results contribute to the growing discussion around potential economic asymmetries in the relationship between the people who create data and intelligent technologies that rely on this data. Our findings also have implications for Wikipedia research on content coverage and for methods in search auditing and UGC research.

## Acknowledgements

This work was supported in part by NSF grants 1707296 and 1815507. We would like to thank our anonymous reviewers for their feedback and assistance in improving the manuscript.

This version includes a bibliography entry that was missing from the first version of the text.